\documentclass[10pt,conference]{IEEEtran}
\IEEEoverridecommandlockouts
\usepackage{cite}
\usepackage{amsmath,amssymb,amsfonts}
\usepackage{hyperref}
\usepackage{algorithm}
\usepackage{algorithmic}
\usepackage{enumitem}
\usepackage{graphicx}
\usepackage{tikz}
\usetikzlibrary{calc}
\usetikzlibrary{positioning, shapes, arrows, arrows.meta, fit,backgrounds}
\usepackage{textcomp}
\usepackage{xcolor}
\usepackage{multirow}
\usepackage{subfig}
\usepackage{comment}
\usepackage{amsthm}  
\usepackage{amsmath} 
\usepackage{amssymb} 
\usepackage{svg}

\theoremstyle{definition}

\theoremstyle{remark}

\ifcsname proof\endcsname\else
  
  \providecommand{\proofname}{Proof}
\fi

\usepackage[font={small}]{caption}

\def\BibTeX{{\rm B\kern-.05em{\sc i\kern-.025em b}\kern-.08em
    T\kern-.1667em\lower.7ex\hbox{E}\kern-.125emX}}

\usepackage{listings}
\usepackage{xcolor} 

\definecolor{lightgray}{rgb}{0.95, 0.95, 0.95}

\lstdefinelanguage{json}{
    basicstyle=\normalfont\ttfamily,
    comment=[l]{//},
    morestring=[b]",
    showstringspaces=false,
    morekeywords={:},
    alsoletter={:}
}

\begin{document}

\title{Video Quality Monitoring \\ for Remote Autonomous Vehicle Control%
\thanks{This research has been supported by impact-xG, an open call project funded by the TARGET-X project. TARGET-X is co-funded by the European Union through the Smart Networks and Services Joint Undertaking (SNS JU) under Horizon Europe research and innovation programme under Grant Agreement No 101096614. Views and opinions expressed are however those of the authors only and do not necessarily reflects those of TARGET-X, the European Union, or European Commission. Neither the European Union nor the granting authority can be held responsible for them.}}

\author{
\IEEEauthorblockN{
Dimitrios Kafetzis\textsuperscript{1},
Nikos Fotiou\textsuperscript{1,2},
Savvas Argyropoulos\textsuperscript{2}, Jad Nasreddine\textsuperscript{3} and
Iordanis Koutsopoulos\textsuperscript{1}}
\IEEEauthorblockA{\textsuperscript{1}Athens University of Economics and Business, Department of Informatics, Athens, Greece\\
\textsuperscript{2}StreamOwl, Athens, Greece\\
\textsuperscript{3}i2CAT Foundation, Barcelona, Spain}
}
\maketitle

\begin{abstract}
The delivery of high-quality, low-latency video streams is critical for remote autonomous vehicle control, where operators must intervene in real time. However, reliable video delivery over Fourth/Fifth-Generation (4G/5G) mobile networks is challenging due to signal variability, mobility-induced handovers, and transient congestion. In this paper, we present a comprehensive blueprint for an integrated video quality monitoring system, tailored to remote autonomous vehicle operation. Our proposed system includes subsystems for data collection onboard the vehicle, video capture and compression, data transmission to edge servers, real-time streaming data management, Artificial Intelligence (AI) model deployment and inference execution, and proactive decision-making based on predicted video quality. The AI models are trained on a hybrid dataset that combines field-trial measurements with synthetic stress segments and covers Long Short-Term Memory (LSTM), Gated Recurrent Unit (GRU), and encoder-only Transformer architectures. As a proof of concept, we benchmark 20 variants from these model classes together with feed-forward Deep Neural Network (DNN) and linear-regression baselines, reporting accuracy and inference latency. Finally, we study the trade-offs between onboard and edge-based inference. We further discuss the use of explainable AI techniques to enhance transparency and accountability during critical remote-control interventions. Our proactive approach to network adaptation and Quality of Experience (QoE) monitoring aims to enhance remote vehicle operation over next-generation wireless networks.
\end{abstract}


\section{Introduction}
\label{sec:introduction}

Autonomous driving systems are categorized into different levels based on their degree of automation, ranging from Level 2, which involves partial automation with human supervision, to Level 5, representing full driving autonomy \cite{sae_j3016}. Level 2 autonomy is typically realized with remote human operators who monitor live video feeds from vehicles and intervene when necessary. For instance, Nissan is currently testing remote teleoperation technology, employing human operators at remote locations who can assume control if automated systems encounter issues \cite{nissan_remote_tech}. In such scenarios, reliable, high-quality video streams are critical, as video quality degradation directly impacts real-time decision-making and compromises safety.

Recent surveys confirm the feasibility of teleoperation over Fourth/Fifth-Generation (4G/5G) mobile networks but also highlight the necessity for robust latency and bandwidth guarantees \cite{Amador2022SurveyRORV}. Maintaining high Quality of Experience (QoE) for vehicular video streaming is challenging due to mobility-induced radio handovers, frequent packet loss, and varying signal quality \cite{Quadros2016QoE_Vehicular}. Traditional network metrics (e.g. throughput, latency) inadequately reflect user-perceived quality, prompting perceptual metrics such as Video Multi-Method Assessment Fusion (VMAF), which strongly correlate with human opinions \cite{Rassool2017VMAF}.

Addressing these challenges requires an integrated, end-to-end video quality monitoring system encompassing data collection onboard the vehicle, video capture and compression, data transmission, real-time data processing, and proactive decision-making. Within this pipeline, Artificial Intelligence (AI)-based QoE prediction models serve as critical tools for proactive network adaptation. Recent advances in Deep Learning (DL), particularly Recurrent Neural Networks (RNNs) such as Long Short-Term Memory (LSTM) \cite{hochreiter1997lstm}, Gated Recurrent Units (GRU) \cite{cho2014rnnencdec}, and self-attention-based Transformers \cite{vaswani2017attention}, have demonstrated strong capabilities for short-term QoE forecasting \cite{Ahmad2022QoEPredictionTutorial}.

In this paper, we present a blueprint for an integrated video quality monitoring system tailored specifically for vehicular teleoperation scenarios. Our contributions include:
\begin{itemize}
    \item A complete system architecture for collecting, transmitting, processing, and managing real-time video quality data for teleoperated vehicles.
    \item A methodology for constructing hybrid dataset combining real-world measurements with synthetic data for realistic evaluation scenarios.
    \item A comparison of 20 model variants—LSTM, GRU, Transformer, Deep Neural Network (DNN), and linear regressor baselines—to pinpoint the fastest, most accurate short-term QoE forecaster.
    \item A discussion on deployment considerations, including onboard vs. edge-based inference and the incorporation of explainable AI techniques for decision transparency.
\end{itemize}

The remainder of the paper is organized as follows. Section~\ref{sec:related} reviews relevant work in vehicular teleoperation and QoE assessment. We detail the proposed end-to-end system architecture in Section~\ref{sec:system}. Section~\ref{sec:aimodel} outlines our dataset generation and deep learning components, while Section~\ref{sec:inference} addresses real-time inference workflows. Section~\ref{sec:explainability} discusses strategies for explainability in diagnosing QoE degradations, and Section~\ref{sec:conclusion} concludes the paper.

\section{Related Work}
\label{sec:related}

\paragraph{Teleoperated Driving Over 4G/5G Networks}
Teleoperated driving, in which a remote human operator can control a vehicle in real time, heavily relies on high-throughput and low-latency wireless links. Early studies established the feasibility of teleoperation via 4G networks, but also highlighted practical limitations such as fluctuating uplink throughput and delayed handovers \cite{Neumeier2019TMA}. With the rollout of 5G, multiple pilot systems have demonstrated improved coverage, lower latency, and more consistent quality for teleoperated driving \cite{Kakkavas2022TeSo5G}. Nonetheless, large-scale deployment must address factors like dedicated uplink capacity, edge computing to minimize round-trip times, and network slicing to ensure reliability \cite{LucasEstan2023VTC}.

\paragraph{QoE Models for Vehicular Video Streaming}
Vehicle mobility induces variable radio conditions (e.g., handovers, received signal fluctuations), making it difficult to maintain stable video quality. Traditional network-layer 
Quality of Service (QoS) metrics (packet loss, delay) cannot fully capture user-perceived QoE under these rapid fluctuations. Several works propose QoE-driven strategies that combine network-level adaptation with application-level buffering and rate control \cite{Erfanian2022QoCoVi}. They often employ distributed caching or dynamic path selection to address bandwidth variability. Surveys on video streaming in vehicular ad-hoc networks (VANETs) confirm the importance of leveraging QoE estimates to guide route selection and reduce rebuffering \cite{Zribi2019VANETvideo}.

\paragraph{AI-Based QoE Prediction}
Deep learning methods are increasingly used to predict QoE from time-series data of network measurements and video application logs. RNNs, including LSTM and GRU, have shown particular promise due to their ability to capture temporal relationships in streaming data \cite{Eswara2020LSTMQoE}. Through short-term QoE forecasting such models enable proactive decisions that mitigate impending video stalls or drops in resolution. In parallel, Transformer-based architectures, originally devised for natural language processing, are being investigated for video quality assessment tasks. Transformers rely on a self-attention mechanism, which allows the model to weigh the importance of data at different points in time and directly relate them, capturing both immediate effects (e.g., sudden network jitter) and longer-term influences (e.g., gradual buffer depletion), making them effective when disruptions and impairments span multiple time scales \cite{Kossi2024TransformerVQA}.

\paragraph{Video Quality Assessment Metrics}
Developers and researchers rely on objective metrics to measure the perceptual quality of encoded video. Although early metrics like PSNR remain widespread, they do not always correlate strongly with user perception \cite{Wang2004SSIM}. In contrast, the VMAF metric combines multiple perceptual video quality features, such as visual fidelity, temporal smoothness, and distortion measures, through a machine learning model trained on viewer opinions to accurately reflect subjective mean opinion scores. Validations have reported high correlations between VMAF and viewer ratings for diverse video content and encoding conditions \cite{Rassool2017VMAF}. Consequently, VMAF is often used to benchmark adaptive streaming systems and train AI-based QoE prediction models.

In the following sections, we build upon these insights to present an end-to-end pipeline for teleoperated vehicles—data collection, video processing, network adaptation, and QoE prediction—so that perceived quality stays high in real-world conditions.

\section{System Architecture}
\label{sec:system}

\begin{figure*}[!t]
    \centering
    \includegraphics[width=0.92\linewidth]{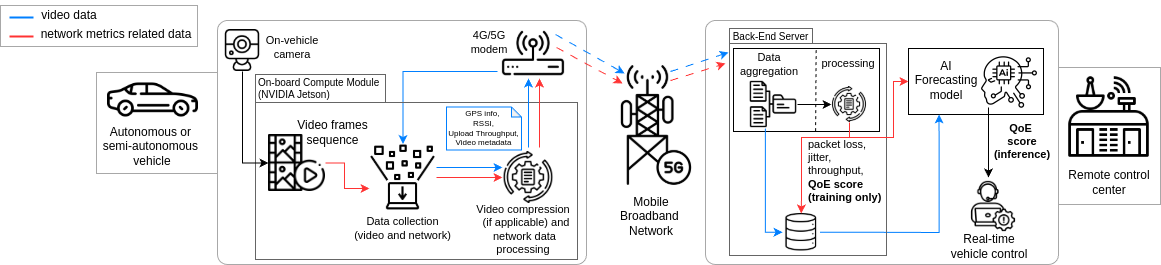}
    \caption{High-level architecture of the proposed remote driving system.}
    \label{fig:system_arch}
\end{figure*}

This section outlines our end-to-end teleoperation stack—video capture, encoding, transmission, and quality monitoring. Figure \ref{fig:system_arch} groups the setup into three layers: in-vehicle hardware, the wireless network, and the remote control station.

\subsection{Onboard Vehicle Hardware and Modem Integration}
The teleoperated vehicle is equipped with a camera that capture a frontal view in high-definition. High-resolution video feeds from vehicles generate significant uplink traffic; thus, efficient compression is essential to minimize bandwidth usage and maintain consistent QoE. Depending on the available resources, the captured video frames can be either compressed onboard using hardware-accelerated encoders (e.g., H.264/Advanced Video Coding (AVC) or H.265/High Efficiency Video Coding (HEVC)) or transmitted raw to an edge node for compression. The selection between onboard and edge compression is influenced by central processing unit (CPU) or graphics processing unit (GPU) resource availability, power constraints, and real-time network conditions. In addition, the vehicle hosts a 5G smart phone acting as an access point for connectivity to the mobile broadband network. A mobile application periodically reports radio signal-quality indicators such as Reference Signal Received Power (RSRP) and Reference Signal Received Quality (RSRQ). These radio-level metrics are timestamped and correlated with video quality metrics, enabling comprehensive QoE assessments.

\subsection{Video Capture and Compression}
Offloading raw video frames to an edge node for real-time compression can leverage the more powerful cloud or edge servers, thereby reducing the in-vehicle computational footprint; however, it requires higher instantaneous throughput. In contrast, local encoding offers bandwidth savings, especially beneficial when uplink capacity is constrained. Specifically, if the network capacity is oftentimes limited (e.g., in urban canyons with poor coverage), compressing video onboard to transmit a reduced amount of data becomes crucial for sustaining acceptable QoE \cite{Zribi2019VANETvideo,Erfanian2022QoCoVi}. However, onboard processing may introduce additional latency and increase power consumption.

Another subtle but critical trade-off concerns error resilience. While uncompressed video is bandwidth-intensive, every frame is independent, making the stream relatively tolerant to bit or packet losses. In contrast, compressed video relies on predictive encoding (e.g., reference frames), so even minor errors can cascade, causing visible artifacts (e.g., blockiness, smearing) in subsequent frames. In our tests, we observed marked quality drops in compressed streams even with moderate packet loss. Consequently, operators must weigh the benefits of reduced data volume against the risk of amplified degradation in challenging radio environments.

\subsection{Performance Measurement and QoE Monitoring}
Our system incorporates a \emph{measurement module} that observes network-layer metrics (packet loss, latency, jitter) and application-layer QoE signals (e.g., resolution switching, rebuffering). This module collects the following data:
\begin{itemize}
    \item \emph{Network Metrics:} Packet loss rate, jitter, throughput, radio-level metrics (RSRP, RSRQ).
    \item \emph{Application Events:} Video stalls, resolution changes, or frame drops from the encoder.
    \item \emph{QoE Scores:} Obtained via a perceptual metric (e.g., a VMAF-like estimator).
\end{itemize}

VMAF is a full-reference objective metric: it receives a pair of streams—the reference and its encoded counterpart—extracts several perceptual indicators and feeds them to a regression model trained on viewer studies, outputting a single \(0\text{–}100\) quality score that correlates tightly with Mean-Opinion-Scores (MOS) \cite{Rassool2017VMAF}. 
In our pipeline this score acts both as the ground-truth label for training the QoE forecaster and as an indicator for operator-perceived quality.  

These measurements are periodically batched and transmitted to a back-end server, where they can be aggregated and pre-processed for real-time or offline analysis.

\subsection{Remote Operator and Control Station}
At the other end of the pipeline, a remote control station (or teleoperation cockpit) receives the video feed and issues commands back to the vehicle. In typical setups, the downlink requirements (for control signals) are much smaller than the uplink capacity needed for continuous high-definition (HD) or ultra-high-definition (UHD) video \cite{Amador2022SurveyRORV}. Nonetheless, low downlink latency is vital for ensuring timely actuation. Consequently, the control station might employ multi-homed or redundant network interfaces (e.g., wired + 5G) to guard against outages. Additionally, edge nodes near the base station may host AI components that assist with real-time video processing or QoE prediction.

\subsection{Summary of Key Design Considerations}
\begin{itemize}
    \item Balancing the trade-off between local compression (reducing uplink data volume but introducing greater CPU overhead and error sensitivity) and uncompressed video (requiring higher throughput but offering stronger resilience to network errors).
    \item Dynamically adjusting bitrate or resolution in response to changing radio conditions to maintain QoE.
    \item Continuously collecting multi-layer metrics (network, application, radio) with minimal overhead to enable QoE monitoring and forecasting.
    \item Scaling to multiple vehicles in dense urban scenarios without overwhelming either the wireless network or the edge infrastructure.
\end{itemize}


\section{AI Model: Dataset Construction, Architecture and Evaluation}
\label{sec:aimodel}

This section explains how the training dataset was built, which models were benchmarked, and how their performance was assessed.  
It begins with the hybrid dataset and the preprocessing procedure.
We list the candidate model classes—linear regressors, feed-forward DNNs, gated-recurrent networks (LSTM, GRU) and encoder-only Transformers—and close with the accuracy and latency metrics.

\subsection{Data Collection and Labeling}
\label{subsec:data_collection}

\paragraph{Real-World Vehicular Measurements}
We collected real traces from a vehicle driving in an urban environment streaming video to an end-host over a mobile network, similar to the procedure in \cite{Neumeier2019TMA}. The receiving host continuously recorded:
\begin{itemize}
    \item Packet loss rate, jitter, and throughput, derived from captured network traffic.
    \item Global Positioning System (GPS)-based vehicle speed (km/h), affecting the tolerance of the streamed video to network errors.
    \item A VMAF-like perceptual score, computed at intervals on a low-latency edge node to estimate subjective video quality.
\end{itemize}
Each metric was timestamped to facilitate window-based aggregation.

\paragraph{Synthetic Data Generation}
While real data capture authentic mobility and wireless fluctuations, some conditions (e.g., extreme packet loss) were underrepresented. To address these coverage gaps, we generated synthetic records using network emulation tools available in the Linux kernel (e.g., \texttt{tc-netem}) to artificially introduce controlled packet loss, latency spikes, and throughput drops, following guidelines from \cite{Erfanian2022QoCoVi}, which recommends incorporating boundary conditions (e.g., very poor signal or abrupt throughput drops). Each synthetic record spans a short time window (e.g., 10\,s), containing:
\textit{packet\_loss\_rate} in the range \(0\%\) to \(5\%\), \textit{jitter} (10--100\,ms), and \textit{throughput} (5--50\,Mbps).

\subsection{Feature Engineering and Sampling}
\label{subsec:feature_engineering}
To transform raw logs into time-series inputs, we segmented each trace into fixed-length windows (e.g., 10\,s). For each window, we calculate the VMAF score. We normalize all features (loss, jitter, throughput, speed, etc.) to the \([0,1]\) range using min-max scaling. Each input sequence thus becomes a 2D array of size: $L \times (\text{number of features})$, where \(L\) is the sequence length (e.g., 5 steps of 2\,s each). Samples are created with a sliding context of five consecutive windows (50 sec of history).  
The first 80 \% of the timeline is allocated to training and validation, while the remaining 20 \% is kept as an unseen test set, ensuring zero information leakage.

\subsection{Model Classes}
\label{subsec:model_families}
To understand how architectural complexity and regularisation affect both prediction error and inference speed, we organised candidate AI models into five model classes that span the range from memory‑less linear regressors to modern attention mechanisms.  Inside each class we varied only one design axis at a time—hidden-state width, network depth, or dropout rate—so that we could isolate the effect of extra capacity (i.e., more trainable parameters) and stronger regularisation (i.e., greater resistance to over-fitting). The 5 classes and the variants within each class, are as follows:

\begin{enumerate}[label=(\alph*)]
\item LSTM recurrent networks:
      \begin{itemize}
        \item \emph{Basic LSTM} – one LSTM layer with 32 hidden units followed by a 128‑unit self‑attention block that re‑weights the five hidden‑state vectors before the output.
        \item \emph{Wide LSTM} – widens the single layer to 100 units to test whether a broader state vector improves accuracy on short sequences.
        \item \emph{Deep LSTM} – three stacked LSTM(32) layers with dropout 0.2 between them, assessing whether additional depth yields lower error once over‑fitting is controlled.
      \end{itemize}

\item GRU recurrent networks:
      \begin{itemize}
        \item \emph{Basic GRU} – one GRU layer with 32 hidden units plus the same 128‑unit self‑attention block.
        \item \emph{Wide GRU} – increases hidden units to 64, adding capacity with minimal latency overhead.
        \item \emph{Deep GRU} – three GRU(32) layers separated by dropout 0.20 to gauge the value of extra recurrence depth.
      \end{itemize}

\item Encoder‑only Transformers:
      \begin{itemize}
        \item \emph{Basic Transformer} – one encoder block with 2 heads, feed‑forward dimension 64, dropout 0.10.
        \item \emph{Four‑heads} – raises the head count to 4 to capture richer pairwise interactions at the same depth.
        \item \emph{Larger‑ff} – doubles the feed‑forward dimension to 128, testing the impact of a wider projection layer.
        \item \emph{Low‑drop} – keeps the Basic layout but reduces dropout to 0.05, probing sensitivity to regularisation strength.
      \end{itemize}

\item Feed‑forward DNNs:
      \begin{itemize}
        \item \emph{Basic DNN} – two dense layers \([64, 32]\) with ReLU activations \cite{nair2010relu} and dropout 0.2.
        \item \emph{Deep DNN} – three layers \([128, 64, 32]\) to see whether extra depth benefits a purely feed‑forward approach.
        \item \emph{ELU DNN} – identical to Basic but with ELU activations \cite{clevert2016elu}, isolating the effect of the activation function.
        \item \emph{High‑drop DNN} – same topology as Basic DNN but with dropout 0.4 for stronger regularisation against over‑fitting.
      \end{itemize}

\item Linear regressors:
      \begin{itemize}
        \item \emph{Basic Linear} – ordinary least squares on the flattened sequence of five windows $\times$ \(d\) features (\(d=6\)), i.e.\ 30 input scalars, ignoring temporal order.
        \item \emph{L1 (Lasso)} \cite{tibshirani1996lasso} – adds an \(L_{1}\) penalty (\(\lambda=0.01\)) to encourage sparsity.
        \item \emph{L2 (Ridge)} \cite{hoerl1970ridge} – applies an \(L_{2}\) penalty (\(\lambda=0.01\)) to shrink weights smoothly.
        \item \emph{ElasticNet} \cite{zou2005elasticnet} – combines \(L_{1}=L_{2}=0.005\) for a balance between sparsity and stability.
      \end{itemize}

\end{enumerate}

All recurrent models employ sequence length \(\ell=5\) (50 sec of telemetry) and attach the same attention layer so their outputs remain comparable.  Hyper-parameters shared across model classes (learning rate, optimiser, early-stopping criteria) follow the training methodology in Section\,\ref{subsec:training_experiments}.  This catalogue provides a transparent ladder of complexity against which later evaluation results in Table\,\ref{tab:test_results} and Figure \ref{fig:kde_abs} can be interpreted. All preprocessing, training, and evaluation code for every variant is open-sourced at GitHub \cite{qoe_repo}.

\subsection{Model Training and Evaluation}
\label{subsec:training_evaluation}

\subsubsection*{Training methodology}
\label{subsec:training_experiments}
In our implementation the linear regressor is trained by minimising the mean-squared-error (MSE) objective \cite{hyndman2006accuracy}, whereas the neural networks are trained by minimising the Keras’s Log-Cosh loss \cite{keras_docs}.  We further use Keras’s \texttt{EarlyStopping} and \texttt{ReduceLROnPlateau} callbacks to halt training or lower the learning rate when validation improvement stalls.
Hyper-parameters—layer width, depth, dropout and optimiser schedule—are tuned through the Keras Hyperband tuner.

\subsubsection*{Evaluation methodology}
Final performance is measured on the chronologically held-out test set with three metrics:
\begin{itemize}
\item Mean Absolute Error (MAE) \cite{hyndman2006accuracy} – easy to interpret in the original VMAF scale.
\item Root-Mean-Squared Error (RMSE) \cite{hyndman2006accuracy} – emphasises large mistakes.
\item Inference latency – average GPU time per sample (batch\,=\,16) on an RTX 2060 SUPER. 
\end{itemize}
All models are benchmarked under identical software (Python 3, TensorFlow 2.10, CUDA 11.8) and hardware (Intel i7-9700K, RTX 2060 SUPER, 32 GB RAM) conditions.

\subsubsection*{Evaluation Results}

\begin{figure}[t]
  \centering
  \includegraphics[width=.93\linewidth]{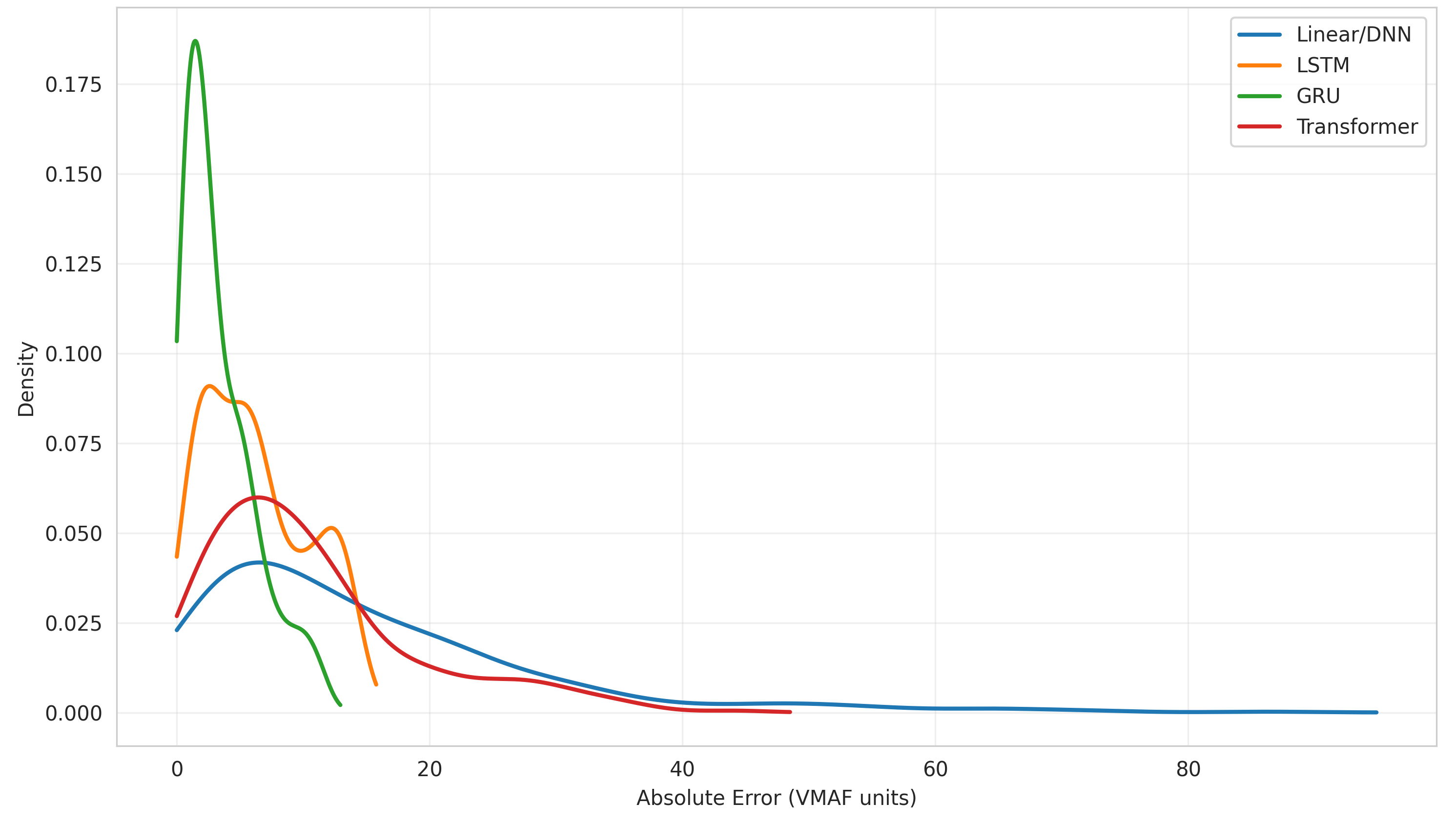}
  \caption{Absolute-error density per model class.  GRU (green) peaks sharply at zero, LSTM and Transformer are wider, Linear/DNN shows the longest tail.}
  \label{fig:kde_abs}
\end{figure}

Figure \ref{fig:kde_abs} and Table\,\ref{tab:test_results} corroborate the numerical ranking:  
the GRU model class not only achieves the best mean errors but also the tightest error spread, whereas LSTM and Transformer exhibit broader distributions and the Linear/DNN baselines trail at the high-error end.  
We conclude that a single-layer Basic GRU offers the best accuracy–latency balance for proactive QoE control in remote driving.

\begin{table}[h]
\caption{Top-5 Video QoE forecasting model configurations ranked by RMSE and MAE (measured in VMAF units).}
\centering
\begin{tabular}{lccc}
\hline
\noalign{\vskip 0.4ex}             
Model variant & RMSE & MAE & Latency (ms)\\
\noalign{\vskip 0.1ex}             
\hline
\noalign{\vskip 0.7ex} 
Basic GRU (32)   & 1.62 & 1.41 & 66 \\
Wide GRU (64)    & 2.85 & 2.40 & 59 \\
Basic LSTM (32)  & 4.47 & 3.87 & 66 \\
Deep GRU (3×32)  & 4.61 & 4.08 & 68 \\
Deep LSTM (3×32) & 6.71 & 5.81 & 67 \\
\noalign{\vskip 0.4ex} 
\hline
\end{tabular}
\label{tab:test_results}
\end{table}

A single-layer \emph{Basic GRU} with 32 hidden units produces the smallest errors—RMSE = 1.62 and MAE = 1.41—equivalent to a deviation of \(<\!1.6\,\%\) of the full VMAF scale and practically imperceptible to operators.  
Every GRU variant outperforms its LSTM counterpart of equal width or depth; the best encoder-only Transformer ranks just below the Wide GRU, confirming that self-attention needs a larger dataset to rival gated recurrence under the present data regime.

Latency remains comfortably within operational bounds.  
On the RTX 2060 SUPER the Basic GRU averages 66 msec of inference time (batch = 16); adding 18 msec for on-vehicle feature extraction, two 20 msec 5G hops (uplink and return), and 7 msec for UI rendering gives an end-to-end turnaround of about 131 msec.  
Because the model forecasts the QoE two seconds ahead, the control loop still has roughly \(2.0-0.13 \approx 1.9\) sec to react—well inside the sub-150 msec budget often cited for safe tele-operation. 

\subsubsection*{Discussion}
GRUs achieve the best results because their two gates keep only the recent context needed for a ten-second forecast and discard older, less relevant data. LSTMs carry extra internal state that adds complexity without boosting accuracy for this medium-range horizon, while Transformers must enlarge their attention heads and feed-forward layers to compete, increasing parameter count and latency.  
Linear regressor and shallow DNN baselines finish inference in about 55 msec but their prediction errors are two-to-four times larger.  
Transformers approach GRU accuracy only after widening, which lifts latency beyond 60 msec yet still cannot match the GRU’s MAE. With RMSE equal to 1.6 VMAF units and a 66 msec inference time, the single-layer 32-unit \emph{Basic GRU} offers the best accuracy–latency balance for live QoE forecasting.  
A wider 64-unit GRU shaves latency to 59 msec at the cost of a modest error increase, whereas adding more layers to any recurrent model raises computation without meaningful accuracy gains.

\section{Inference and Real-Time Deployment}
\label{sec:inference}

After training the DNN models, we integrate them into the teleoperation system for short-term QoE forecasts, outlining where inference runs—onboard or at the edge—and how the results guide the control loop.

\subsection{Onboard vs.\ Edge-Based Inference}
\label{subsec:onboard_vs_edge}
\paragraph{Onboard Inference}
Running the QoE prediction model directly on the vehicle’s embedded hardware (e.g., an Advanced RISC Machines (ARM)-based system-on-chip (SoC) with GPU acceleration) enables low-latency decision-making without reliance on upstream connectivity. This is especially beneficial if timely adaptive measures (e.g., adjusting the video encoder or requesting more uplink resources) must be taken whenever a QoE drop is imminent. However, onboard inference increases processing overhead and may contend with other safety-critical tasks for CPU/GPU resources \cite{Vu2022GRU-QoE}.

\paragraph{Edge-Based Inference}
Alternatively, the computation for the inference using a model can be offloaded to a nearby edge server or cloud platform. This reduces the computational burden in the vehicle but requires reliable connectivity to transmit relevant measurement features (e.g., packet loss, throughput, speed) in near-real time. Although 5G networks promise low-latency links, fluctuations in coverage or backhaul congestion can still impact how quickly the edge server receives data and returns predictions. For teleoperated driving, any additional round-trip latency must be carefully weighed against the potential gains in inference accuracy from running on powerful hardware \cite{LucasEstan2023VTC}.

\subsection{Real-Time Pipeline}
\label{subsec:realtime_pipeline}
The typical inference pipeline steps when using short sliding windows of data are following:
\begin{enumerate}[label=(\alph*)]
    \item \emph{Data Buffering:} Recent network and vehicular metrics are cached for a window of length \(L\) seconds (e.g., 5--10\,s).
    \item \emph{Feature Preprocessing:} Raw measurements (packet loss, jitter, throughput, speed) are scaled according to the min-max statistics from training.
    \item \emph{Model Inference:} The LSTM/GRU/Transformer model predicts the QoE for the next interval (\(t + \Delta\)), using the last \(L\) seconds as input.
    \item \emph{Feedback Loop:} Predictions are fed into adaptive logic that may adjust the encoding bitrate/resolution onboard. Also to trigger proactive buffer management or packet prioritization. As well as to alert the remote operator if a substantial QoE drop is imminent.
\end{enumerate}

\subsection{Overhead and Latency}
\label{subsec:overhead_latency}
Inference latency depends on factors like:
\begin{itemize}
    \item \emph{Model Complexity:} Transformers tend to have larger parameter counts, though for shorter sequences (e.g., \(L \leq 10\)), an optimized implementation can still run in a few milliseconds on a GPU or even an embedded accelerator.
    \item \emph{Compute Platform:} Onboard hardware (NVIDIA Jetson) may have limited Compute Unified Device Architecture (CUDA) cores or memory. While an edge server might have more resources, processing multiple vehicles or other concurrent services can introduce congestion, potentially increasing inference latency.
    \item \emph{Data Transfer Time:} For edge-based inference, the overhead of sending telemetry to the server and receiving predictions can add 5--20\,msec in 5G networks, assuming moderate load \cite{Amador2022SurveyRORV}.
\end{itemize}
In general, LSTM/GRU architectures incur lower computational cost than Transformers of similar capacity \cite{Eswara2020LSTMQoE}, making them appealing for in-vehicle deployment. Yet, Transformers could yield superior performance when capturing extended temporal patterns, such as prolonged radio degradation or persistent vehicular congestion, justifying their higher overhead.

Overall, these design considerations highlight the trade-offs between onboard and edge inference, the importance of low-latency feedback for teleoperation, and the need for lightweight yet accurate QoE forecasting models. In the next section, we discuss potential strategies for interpreting the model’s predictions through explainable AI approaches, which can help diagnose the root causes of sudden QoE drops.

\section{Incorporating Explainability in Video QoE prediction}
\label{sec:explainability}

Although DNNs have demonstrated strong predictive performance for QoE forecasting, their opaque nature poses challenges for deploying them in mission-critical teleoperation systems. Network operators, automotive original equipment manufacturers (OEMs), and end-users benefit from understanding why a model predicts an impending quality drop—particularly if corrective actions must be taken urgently \cite{Kossi2024TransformerVQA}.

\subsection{Motivation for Explainable AI}
Explainable AI (XAI) methods can shed light on key factors influencing predicted QoE scores, providing accountability for decisions taken due to video quality degradation. For example, if a sudden QoE deterioration triggers a human operator to take immediate control of the vehicle by applying brakes, an XAI method could clearly communicate to the operator that the decision was prompted by a sudden increase in packet loss or jitter. Benefits include:
\begin{itemize}
    \item \emph{Trust and Transparency:} Stakeholders gain confidence in system decisions by understanding how network and vehicular parameters impact QoE.
    \item \emph{Debugging and Network Optimization:} Operators prioritize corrective actions by identifying root causes like poor signal strength or jitter spikes.
    \item \emph{Safety and Compliance:} Explainability ensures regulatory compliance and accountability in safety-critical scenarios.
\end{itemize}

\subsection{Approaches for Explainability}
\paragraph{Feature Attribution Methods}
Common post-hoc techniques such as Integrated Gradients can highlight which input features (e.g., \textit{throughput}, \textit{jitter}, \textit{speed}) most contributed to a specific QoE prediction. For LSTM or GRU models, these methods track gradient flows through recurrent cells, producing a temporal relevance map \cite{Dinaki2021DeepQoEForecast}. 

\paragraph{Attention Weights in Transformers}
For Transformer-based architectures, the self-attention mechanism inherently provides attention weights that indicate how each time-step or feature dimension “attends to” others \cite{Kossi2024TransformerVQA}. The weights show which earlier readings (e.g., a packet-loss spike) most influenced the QoE forecast. Caution is advised, however, as raw attention weights may not perfectly align with causal explanations \cite{Ahmad2022QoEPredictionTutorial}.

\paragraph{Local Surrogate Models}
A simple surrogate (e.g., linear regressor or decision tree) can be fitted around a single prediction; Local Interpretable Model-Agnostic Explanations (LIME) can approximate the DNN’s local decision boundary by perturbing input features and observing changes in QoE output. Although costlier, such local surrogates reveal model behaviour in specific situations (e.g., poor coverage).

\subsection{Practical Considerations}
\paragraph{Computation Overhead}
While post-hoc methods like Integrated Gradients or LIME can run offline for diagnostic purposes, real-time teleoperation might necessitate faster or more lightweight explanations—particularly if operators must respond within milliseconds. Precomputing partial explanations for known conditions (e.g., specific speed ranges or throughput thresholds) can mitigate some overhead.

\paragraph{Integration in Control Loops}
Explainability can inform control-loop logic by identifying which parameter adjustments (e.g., raising the encoding bitrate vs.\ switching cell towers) are likely to yield the greatest QoE improvements. In practice, the system can request explanations only when predicted QoE falls below a threshold, conserving resources for critical events.

\paragraph{Edge vs.\ Onboard Explainability}
If inference is performed at the edge or cloud, explanation generation might also reside there, offloading the in-vehicle resource usage. Conversely, purely onboard solutions that run on embedded hardware might require optimized or simplified XAI techniques to avoid interfering with other safety-critical processes.

Overall, explainable AI fosters confidence in QoE forecasting by clarifying the driving factors behind a model’s predictions.

\section{Conclusion and Open Practical Challenges}
\label{sec:conclusion}

Ensuring reliable, high-quality video for teleoperated driving is critical in urban and highway environments alike, where remote human operators must make timely decisions based on visual information. In this paper, we presented a comprehensive blueprint for an integrated video quality monitoring system tailored specifically for remote autonomous vehicle control. The proposed end-to-end system architecture includes subsystems for onboard data collection, video capture and compression, efficient data transmission, AI-based QoE prediction, and proactive network adaptation. Preliminary results from synthetic datasets indicate promising performance, especially for GRU-based models, demonstrating feasibility within acceptable latency constraints. Deployment considerations, such as onboard vs. edge inference, were also explored.

Several practical challenges remain for effective real-world deployment:
\begin{itemize}
    \item \emph{Real-Time Constraints:} On-vehicle inference competes with other critical automotive workloads, necessitating optimized scheduling or hardware acceleration.
    \item \emph{Video Compression Strategy and Energy Considerations:} Balancing bandwidth, error resilience, and onboard processing is critical. Local video compression reduces uplink volume but can magnify the impact of packet losses; sending uncompressed video mitigates errors but increases transmission energy. Finding an optimal real-time trade-off remains an open research challenge.
    \item \emph{Safety and Liability Considerations:} Teleoperation imposes regulatory requirements for system transparency and explainability, especially during critical events.
\end{itemize}

Tackling these challenges will move teleoperated driving from demos to daily use, enabled by stronger networks, edge computing, and robust AI-based solutions.

\bibliographystyle{ieeetr}
\bibliography{timeseries_models_QoE_ADS}

\begin{thebibliography}{10}

\bibitem{sae_j3016}
{SAE International}, ``Taxonomy and definitions for terms related to driving automation systems for on-road motor vehicles (j3016),'' tech. rep., SAE International, April 2021.

\bibitem{nissan_remote_tech}
Y.~Kageyama, ``Nissan tests driverless cars with remote human operators.'' \href{https://apnews.com/article/driverless-japan-nissan-autonomous-technology-5c12444c3931d1c7a0280789d2b0cba9}{AP News article}, 2023.

\bibitem{Amador2022SurveyRORV}
O.~Amador~Molina, M.~Aramrattana, and A.~Vinel, ``A survey on remote operation of road vehicles,'' {\em IEEE Access}, vol.~10, pp.~130135--130154, 2022.

\bibitem{Quadros2016QoE_Vehicular}
C.~W. Quadros, A.~L. Santos, M.~Gerla, and E.~Cerqueira, ``{QoE}-driven dissemination of real-time videos over vehicular networks,'' {\em Computer Communications}, vol.~91-92, pp.~133--147, 2016.

\bibitem{Rassool2017VMAF}
R.~Rassool, ``{VMAF} reproducibility: Validating a perceptual practical video quality metric,'' in {\em 2017 IEEE Int. Symp. on Broadband Multimedia Systems and Broadcasting (BMSB)}, pp.~1--2, 2017.

\bibitem{hochreiter1997lstm}
S.~Hochreiter and J.~Schmidhuber, ``Long short-term memory,'' {\em Neural Computation}, vol.~9, no.~8, pp.~1735--1780, 1997.

\bibitem{cho2014rnnencdec}
K.~Cho, B.~van Merrienboer, D.~Bahdanau, and Y.~Bengio, ``Learning phrase representations using {RNN} encoder--decoder for statistical machine translation,'' {\em arXiv preprint}, vol.~arXiv:1406.1078, 2014.

\bibitem{vaswani2017attention}
A.~Vaswani, N.~Shazeer, N.~Parmar, J.~Uszkoreit, L.~Jones, A.~N. Gomez, L.~Kaiser, and I.~Polosukhin, ``Attention is all you need,'' in {\em Advances in Neural Information Processing Systems (NIPS)}, vol.~30, pp.~5998--6008, 2017.

\bibitem{Ahmad2022QoEPredictionTutorial}
A.~Ahmad, A.~B. Mansoor, R.~Walshe, J.~Qadir, and A.~Hines, ``Supervised-learning-based {QoE} prediction of video streaming in future networks: A tutorial with comparative study,'' {\em IEEE Communications Surveys \& Tutorials}, vol.~24, no.~2, pp.~1400--1430, 2022.

\bibitem{Neumeier2019TMA}
S.~Neumeier, E.~A. Walelgne, V.~Bajpai, J.~Ott, and C.~Facchi, ``Measuring the feasibility of teleoperated driving in mobile networks,'' in {\em Proc. of the 3rd Network Traffic Measurement and Analysis Conf. (TMA)}, IEEE, 2019.

\bibitem{Kakkavas2022TeSo5G}
G.~Kakkavas, K.~N. Nyarko, C.~Lahoud, D.~Kuehnert, P.~Kueffner, M.~Gabriel, S.~Ehsanfar, K.~Mo\ss{}ner, M.~Diamanti, V.~Karyotis, and S.~Papavassiliou, ``Teleoperated support for remote driving over 5g mobile communications,'' in {\em Proc. of IEEE Int. Mediterranean Conf. on Communications and Networking (MeditCom)}, pp.~469--474, 2022.

\bibitem{LucasEstan2023VTC}
M.~C. Lucas-Estañ, B.~Coll-Perales, M.~I. Khan, S.~S. Avedisov, O.~Altintas, J.~Gozalvez, and M.~Sepulcre, ``Support of teleoperated driving with 5g networks,'' in {\em Proc. of the IEEE 98th Vehicular Technology Conf. (VTC-Fall)}, pp.~1--6, 2023.

\bibitem{Erfanian2022QoCoVi}
A.~Erfanian, F.~Tashtarian, C.~Timmerer, and H.~Hellwagner, ``{QoCoVi}: {QoE}- and cost-aware adaptive video streaming for the internet of vehicles,'' {\em Computer Communications}, vol.~190, pp.~1--9, 2022.

\bibitem{Zribi2019VANETvideo}
N.~Zribi, B.~Alaya, and T.~Moulahi, ``Video streaming in vehicular ad hoc networks: Applications, challenges and techniques,'' in {\em Proc. of the 15th Int. Wireless Communications and Mobile Computing Conf. (IWCMC)}, pp.~1221--1226, 2019.

\bibitem{Eswara2020LSTMQoE}
N.~Eswara, S.~Ashique, P.~Mittal, Y.~Chakalabbi, V.~Perra, S.~Choudhury, and S.~S. Channappayya, ``Streaming video qoe modeling and prediction: A long short-term memory approach,'' {\em IEEE Transactions on Circuits and Systems for Video Technology}, vol.~30, no.~12, pp.~4726--4740, 2020.

\bibitem{Kossi2024TransformerVQA}
K.~Kossi, S.~Coulombe, and C.~Desrosiers, ``No-reference video quality assessment using transformers and attention recurrent networks,'' {\em IEEE Access}, vol.~12, pp.~140671--140684, 2024.

\bibitem{Wang2004SSIM}
Z.~Wang, A.~C. Bovik, H.~R. Sheikh, and E.~P. Simoncelli, ``Image quality assessment: from error visibility to structural similarity,'' {\em IEEE Transactions on Image Processing}, vol.~13, no.~4, pp.~600--612, 2004.

\bibitem{nair2010relu}
V.~Nair and G.~E. Hinton, ``Rectified linear units improve restricted boltzmann machines,'' in {\em Proceedings of the 27th International Conference on Machine Learning (ICML)}, pp.~807--814, Omnipress, 2010.

\bibitem{clevert2016elu}
D.~Clevert, T.~Unterthiner, and S.~Hochreiter, ``Fast and accurate deep network learning by exponential linear units {(ELUs)},'' in {\em International Conference on Learning Representations (ICLR), Workshop Track}, 2016.

\bibitem{tibshirani1996lasso}
R.~Tibshirani, ``Regression shrinkage and selection via the lasso,'' {\em Journal of the Royal Statistical Society: Series B (Methodological)}, vol.~58, no.~1, pp.~267--288, 1996.

\bibitem{hoerl1970ridge}
A.~E. Hoerl and R.~W. Kennard, ``Ridge regression: Biased estimation for nonorthogonal problems,'' {\em Technometrics}, vol.~12, no.~1, pp.~55--67, 1970.

\bibitem{zou2005elasticnet}
H.~Zou and T.~Hastie, ``Regularization and variable selection via the elastic net,'' {\em Journal of the Royal Statistical Society: Series B (Statistical Methodology)}, vol.~67, no.~2, pp.~301--320, 2005.

\bibitem{qoe_repo}
D.~Kafetzis, ``Time-series forecasting neural-network models for video qoe prediction.'' \href{https://github.com/Dimitrios-Kafetzis/Time-series_Forecasting_NN-models_for_Video_QoE_Prediction}{GitHub repository}, 2025.

\bibitem{hyndman2006accuracy}
R.~J. Hyndman and A.~B. Koehler, ``Another look at measures of forecast accuracy,'' {\em International Journal of Forecasting}, vol.~22, no.~4, pp.~679--688, 2006.

\bibitem{keras_docs}
{Keras Team}, {\em {Keras API Documentation}}.
\newblock Keras, 2024.

\bibitem{Vu2022GRU-QoE}
H.~T. Vu, V.-S. Pham, T.~H.~T. Nguyen, and H.-C. Le, ``{QoE-Aware Video Streaming Scheme Utilizing GRU-based Bandwidth Prediction and Adaptive Bitrate Selection for Heterogeneous Mobile Networks},'' in {\em Proc. of the IEEE 9th Int. Conf. on Communications and Electronics (ICCE)}, pp.~126--131, 2022.

\bibitem{Dinaki2021DeepQoEForecast}
H.~E. Dinaki, S.~Shirmohammadi, E.~Janulewicz, and D.~C{\^o}t{\'e}, ``Forecasting video {QoE} with deep learning from multivariate time-series,'' {\em IEEE Open Journal of Signal Processing}, vol.~2, pp.~512--523, 2021.

\end{thebibliography}

\end{document}